\documentclass[twocolumn]{revtex4}
\usepackage[dvips]{graphicx}
\usepackage{amsfonts}
\usepackage{float,color,array}

\begin{document}

\title{Emergent skyrmion-based chiral order in zero-field Heisenberg antiferromagnets on the breathing kagome lattice}

\author{Kazushi Aoyama$^1$ and Hikaru Kawamura$^2$}

\date{\today}

\affiliation{ $^1$Department of Earth and Space Science, Graduate School of Science, Osaka University, Osaka 560-0043, Japan \\
$^2$Molecular Photoscience Research Center, Kobe University, Kobe 657-8501, Japan}

\begin{abstract}
We show that classical Heisenberg antiferromagnets on the breathing kagome lattice can be a platform to realize a zero-field topological order of the scalar spin chirality which can be viewed as a miniature skyrmion crystal (SkX) of discrete form with a small number of spins in its magnetic unit cell. In the model, a third nearest-neighbor (NN) antiferromagnetic interaction along the bond direction $J_3$ and the breathing bond-alternation characterized by the ratio of the NN interaction for large triangles to that for small ones, $J_1'/J_1$, are essential. It is found by means of Monte Carlo simulations that a commensurate triple-${\bf Q}$ state appearing for relatively strong $J_3$ at zero field is the noncoplanar state with the SkX structure in the breathing case of $J_1'/J_1 \neq 1$, while in the uniform case of $J_1'/J_1 =1$, it is a collinear state favored by thermal fluctuations. The origin of this chiral order and experimental implications of our result are also discussed. 
\end{abstract}

\maketitle
In magnetic materials, the scalar spin chirality ${\bf S}_1\cdot({\bf S}_2\times{\bf S}_3)$ defined by three localized spins ${\bf S}_i$ often plays an important role. In particular, when a total chirality summed over the whole system is finite, the underlying noncoplanar spin state can be topologically nontrivial in the sense that the spin structure itself and/or the band structure of coupled electrons are characterized by nonzero integer topological numbers \cite{SkX_review_Nagaosa-Tokura_13, 4sub_Martin_prl_08, 4sub_Akagi_jpsj_10, 12subTopo_Barros_prb_14}. Such a chiral order is known to occur at zero field in two dimensions with its origin being multi-spin interactions \cite{4sub_Momoi_prl_97} or a coupling to conduction electrons \cite{4sub_Martin_prl_08, 4sub_Akagi_jpsj_10, 12subTopo_Barros_prb_14, 4sub_Akagi_prl_12}. In this letter, we theoretically show that a breathing bond-alternation of the lattice serves as another mechanism leading to an exotic zero-field chiral order which can be viewed as a miniature version of a skyrmion-crystal (SkX) topological spin texture. 
  
The SkX is a two-dimensional periodic array of magnetic skyrmions each of which is characterized by an integer topological number corresponding to the total solid angle subtended by all the spins, $n_{\rm sk}=\frac{1}{4\pi}\sum_{i,j,k}\Omega_{ijk}$ \cite{SkX_review_Nagaosa-Tokura_13}. Since the solid angle for three spins $\Omega_{ijk}$ is related with the chirality $\chi_{ijk}={\bf S}_i\cdot({\bf S}_j\times{\bf S}_k)$ via $\Omega_{ijk}=2\tan^{-1}\Big[ \displaystyle{\frac{\chi_{ijk}}{ |{\bf S}_i||{\bf S}_j||{\bf S}_k|+ \sum_{\rm cyclic} {\bf S}_i\cdot{\bf S}_j |{\bf S}_k|  } }\Big] $ \cite{SolidAngle_OOsterom_83}, the SkX can be understood as a topological chiral order. The SkX is usually realized in an applied magnetic field irrespective of whether the Dzaloshinskii-Moriya (DM) interaction is present \cite{SkX_Bogdanov_89, SkX_Yi_09, SkX_Buhrandt_13, MnSi_Muhlbauer_09, MnSi_Neubauer_09, FeCoSi_Yu_10, FeGe_Yu_11, Fefilm_Heinze_11, Cu2OSeO3_Seki_12, CoZnMn_Tokunaga_15, GaV4S8_Kezsmarki_15, GaV4Se8_Fujima_17, GaV4Se8_Bordacs_17, VOSe2O5_Kurumaji_17, AntiSkX_Nayak_17, EuPtSi_Kakihana_18, EuPtSi_Kaneko_19} or not \cite{SkX_Okubo_12, SkX_Leonov_15, SkX_Lin_16, SkX_Hayami_16, SkXimp_Hayami_16, SkX_top2_Ozawa_prl_17, SkX-RKKY_Hayami_17, SkX_Lin_18, SkX-RKKY_Hayami_19, SkX-RKKY_Wang_20, SkX-bondaniso_Hayami_21, SkX-bondaniso_Batista_21, SkX-RKKY_Mitsumoto_21, SkX-phaseshift_Hayami_21, Gd2PdSi3_Kurumaji_19, GdRuAl_Hirschberger_natcom_19, GdRu2Si2_Khanh_20}. Recently, it was pointed out that the SkX can be stable even at zero field due to a coupling to conduction electrons on the triangular lattice \cite{SkX_top2_Ozawa_prl_17, SkX-RKKY_Hayami_17}. 
In view of such a situation, we search for a two-dimensional zero-field chiral order as a candidate for a zero-field SkX. 

Since on the three-dimensional pyrochlore lattice having kagome-lattice layers as a building block, a noncoplanar topological spin texture called the hedgehog lattice \cite{Hedgehog_MFtheory_Binz_prb06, Hedgehog_MFtheory_Park_11, Hedgehog_MCtheory_Yang_16, Hedgehog_transport_Zhang_16, Hedgehog_MCtheory_Okumura_19, MnGe_Kanazawa_16, MnSiGe_Fujishiro_19, SFO_Ishiwata_20} is induced at zero field by the combined effect of the third nearest neighbor (NN) antiferromagnetic interaction along the bond direction $J_3$ and a breathing bond-alternation of the lattice \cite{hedgehog_AK_prb_21}, it is naively expected that in the associated two-dimensional system, i.e., $J_3$-rich antiferromagnets on the breathing kagome lattice, a noncoplanar spin state may possibly be realized at zero field. Inspired by this idea, we consider the $J_1$-$J_3$ classical Heisenberg model on the breathing kagome lattice consisting of an alternating array of corner-sharing small and large triangles.

For the uniform kagome lattice only with the NN antiferromagnetic interaction $J_1$, it is well established that spins do not order even at $T=0$ due to a massive ground-state degeneracy resulting from frustration. By introducing additional interactions, three kinds of 12-sublattice noncoplanar states can be stabilized at zero field. 
Among the three, two are cuboc states induced by further NN interactions, but they are not the SkX, as the total chirality summed over the triangles vanishes \cite{Cuboc_Domenge_prb_05, Cuboc_Domenge_prb_08, Cuboc2}. The rest one is a uniform chiral order induced by the coupling to conduction electrons (see Fig. 4 in Ref. \cite{12subTopo_Barros_prb_14}). Although this chiral order is topologically nontrivial, it appears only for a special electron filling. 
The main finding of this work is that a similar but different 12-sublattice uniform chiral order taking a SkX structure is stabilized by the breathing lattice structure at zero field in the presence of relatively strong $J_3$.

The spin Hamiltonian we consider is given by
\begin{equation}\label{eq:Hamiltonian}
{\cal H} = J_1 \sum_{\langle i,j \rangle_S} {\bf S}_i\cdot{\bf S}_j + J_1' \sum_{\langle i,j \rangle_L} {\bf S}_i\cdot{\bf S}_j + J_3\sum_{\langle \langle  i,j \rangle \rangle } {\bf S}_i\cdot{\bf S}_j,
\end{equation}
where $\langle \rangle_{S(L)}$ denotes the summation over site pairs on small (large) triangles. The ratio between the NN interactions on the small and large triangles $J_1$ and $J_1'$, $J_1'/J_1$,  measures the strength of the breathing bond-alternation. The third NN interaction along the bond direction $J_3$ is fixed to be positive (antiferromagnetic), whereas $J_1$ and $J_1'$ may be positive or negative. For the occurrence of the zero-field chiral order, the signs of $J_1$ and $J_1'$ do not matter as long as antiferromagnetic $J_3$ is sufficiently strong. 

We investigate the ordering properties of the Hamiltonian (\ref{eq:Hamiltonian}) by means of Monte Carlo (MC) simulations. In this work, $2\times 10^5$ MC sweeps are carried out and the first half is discarded for thermalization, where our one MC sweep consists of one Heatbath sweep and successive 10 over-relaxation sweeps. Observations are done at every MC sweep, and the statistical average is taken over 4 independent runs. Total number of spins $N$ is related with a linear system size $L$ via $N=3L^2$. By measuring various physical quantities, we identify low-temperature phases. 
For relatively strong $J_3$, a 12-sublattice triple-${\bf Q}$ state characterized by the three commensurate ordering vectors of ${\bf Q}_1=\frac{\pi}{2a}(-1,-\frac{1}{\sqrt{3}})$,  ${\bf Q}_2=\frac{\pi}{2a}(1,-\frac{1}{\sqrt{3}})$, and ${\bf Q}_3=\frac{\pi}{2a}(0,\frac{2}{\sqrt{3}})$ with side length of each triangle $a$ appears, and it takes the three different spin configurations, collinear, coplanar, and noncoplanar structures, depending on the value of $J_1'/J_1$. The noncoplanar state corresponds to a zero-field SkX state.

\begin{figure}
\begin{center}
\includegraphics[width=\columnwidth]{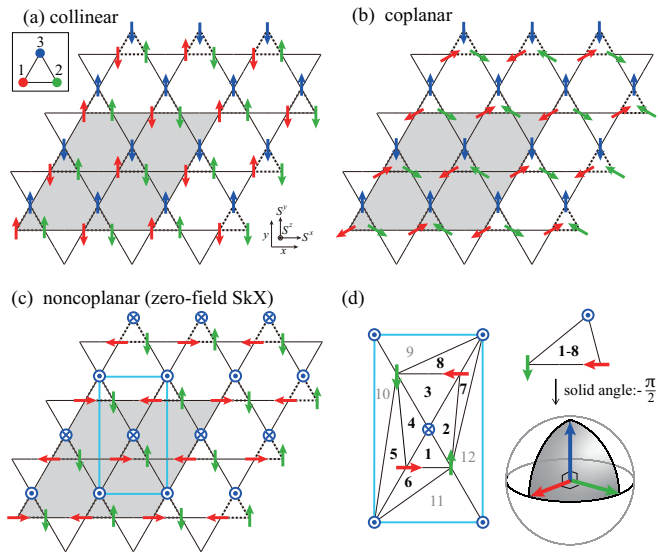}
\caption{Schematically-drawn spin structures of the three triple-${\bf Q}$  states. (a) Collinear, (b) coplanar, and (c) noncoplanar SkX states, where red, blue, and green arrows, respectively, represent spins on the corners of 1, 2, and 3 of the small triangle shown in the inset. 
All the three are 12-sublattice states whose magnetic unit cell is indicated by a gray-colored region. In (c), a cyan rectangle indicates a single skyrmion whose enlarged view is shown in (d). In (d), the solid angle spanned by three spins on each of 1 to 8 triangles is $-\frac{\pi}{2}$, whereas those for 9 to 12 triangles are zero. \label{fig:structure} }
\end{center}
\end{figure}

In the present two-dimensional Heisenberg-spin model, a long-range magnetic order is not allowed at any finite temperature, but the spin correlation develops over more than 500 lattice spacings even just below the transition. Below, we will discuss structures of spins which are not long-range-ordered but correlated over a sufficiently long distance. 

Figure \ref{fig:structure} illustrates the three triple-${\bf Q}$ states, where their elementary unit is the small triangle whose three corners will be called sublattice 1, 2, and 3 [see the inset of Fig. \ref{fig:structure} (a)]. A common feature of the three triple-${\bf Q}$ states is that spins residing on each sublattice, which are represented by the same color arrows in Fig. \ref{fig:structure}, constitute $\uparrow\downarrow\uparrow\downarrow$ chains running along the bond directions, although in reality, $\uparrow$ and $\downarrow$ spins are slightly tilted in the noncollinear states (for details, see Supplemental Material \cite{Suppl}). Suppose that the spin polarization vector of the $\uparrow\downarrow\uparrow\downarrow$ chain on sublattice $\mu$ be $\hat{P}_\mu$. Then, the difference in the three triple-${\bf Q}$ states consists in the relative angles among $\hat{P}_1$, $\hat{P}_2$, and $\hat{P}_3$. In the collinear (coplanar) state shown in Fig. \ref{fig:structure} (a) [(b)], $\hat{P}_1$, $\hat{P}_2$, and $\hat{P}_3$ are in the same direction (plane). In the noncoplanar state shown in Fig. \ref{fig:structure} (c), $\hat{P}_1$, $\hat{P}_2$, and $\hat{P}_3$ are orthogonal to one another, and six spins form a single skyrmion [see a cyan rectangle in Fig. \ref{fig:structure} (c)]; a center spin is pointing down, outer spins are pointing up, and in between, four spins form a vortex. Since it involves only a fixed small number of spins on the discrete lattice sites, it may be called a miniature skyrmion of discrete form, being distinguished from the conventional skyrmion. As one can see from Fig. \ref{fig:structure} (d), the miniature skyrmion is tiled up with twelve triangles among which only inner eight [1-8 in Fig. \ref{fig:structure} (d)] have the nonzero solid angle of $-\frac{\pi}{2}$ because each triangle has orthogonal three spins. Thus, the total solid angle is $-\frac{\pi}{2}\times 8 =-4\pi$ being consistent with the definition of the skyrmion with $n_{\rm sk}=-1$, where the minus sign enters as $\chi_{ijk}$ is defined in the anti-clockwise direction. Since this skyrmion consists of 6 spins, the skyrmion number per 12-sublattice magnetic unit cell is $n_{\rm sk}=\pm2$. We note that on the {\it uniform} kagome lattice, this noncoplanar state is not realized \cite{RMO_Messio_prb_11, RMO-collinear_Grison_prb_20}, but is called a octahedral state in Refs. \cite{RMO_Messio_prb_11, RMO-collinear_Grison_prb_20}. Bearing the above spin structures in our mind, we will turn to the result of our MC simulations.



\begin{figure}
\begin{center}
\includegraphics[width=\columnwidth]{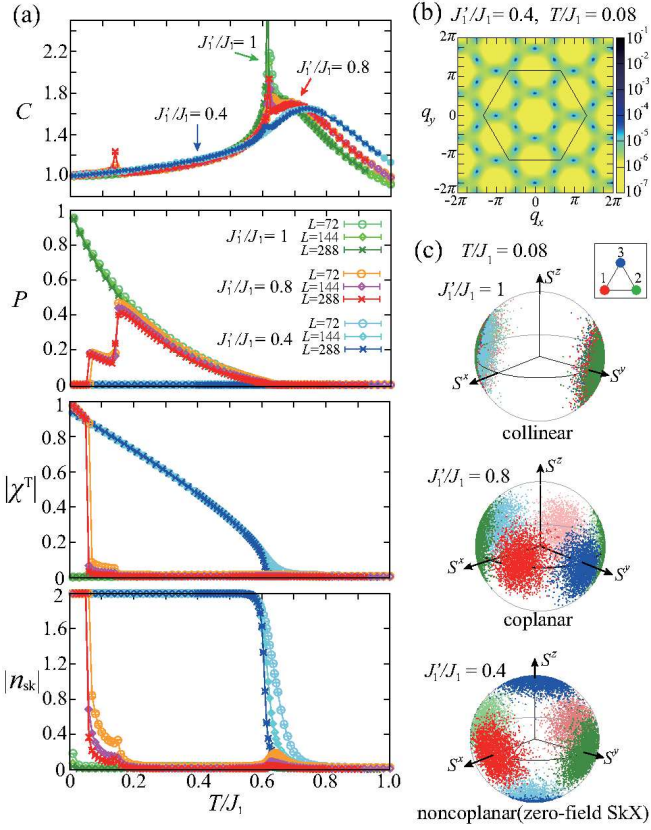}
\caption{MC results obtained for $J_3/J_1=1.2$ with $J_1>0$. (a) Temperature dependence of the specific heat $C$, the spin collinearity $P$, the total scalar chirality $|\chi^{\rm T}|$, and the skyrmion number per magnetic unit cell $|n_{\rm sk}|$ (from top to bottom, respectively), where greenish, reddish, and blueish colored symbols correspond to $J_1'/J_1=1$, $0.8$, and $0.4$, respectively. (b) The spin structure factor $\langle | \frac{1}{N} \sum_i  {\bf S}_i \, e^{i {\bf q}\cdot{\bf r}_i}|^2 \rangle$ obtained in the noncoplanar state at $T/J_1=0.08$ for $J_1'/J_1=0.4$ and $L=72$, where the wave number is measured in units of $1/a$ and a hexagon indicates the border of the Brillouin zone. (c) Spin snapshots mapped onto a unit sphere in the collinear state at $J_1'/J_1=1$ (top), the coplanar state at $J_1'/J_1=0.8$ (middle), and the noncoplanar SkX state at $J_1'/J_1=0.4$ (bottom) all of which are obtained at $T/J_1=0.08$. Red, green, blue dots represent spins on sublattices 1, 2, and 3 shown in the inset. \label{fig:Tdep}   }
\end{center}
\end{figure}

Figure \ref{fig:Tdep} (a) shows the temperature dependence of the specific heat $C$, the spin collinearity $P = \frac{3}{2} \big\langle \frac{1}{N^2}\sum_{i,j} \big( {\bf S}_i\cdot{\bf S}_j\big)^2 - \frac{1}{3} \big\rangle$, the total scalar chirality $|\chi^{\rm T}|=\langle \frac{1}{2L^2}|\sum_{i,j,k \in \bigtriangleup, \bigtriangledown} \chi_{ijk} | \rangle$, and the skyrmion number per magnetic unit cell $|n_{\rm sk}|=\frac{1}{4\pi}\big\langle  \frac{1}{N/12}|\sum'\Omega_{ijk} | \rangle$ for various values of $J_1'/J_1$ at $J_3/J_1=1.2$ with $J_1>0$, where $\langle {\cal O}\rangle$ denotes the thermal average of a physical quantity ${\cal O}$. In $|n_{\rm sk}|$, $\sum '$ denotes the summation over all the triangles that tile up the whole system, and $\Omega_{ijk}$ is evaluated by using spin configurations averaged over 10 MC sweeps to reduce the thermal noise. 

\begin{figure}
\begin{center}
\includegraphics[scale=0.9]{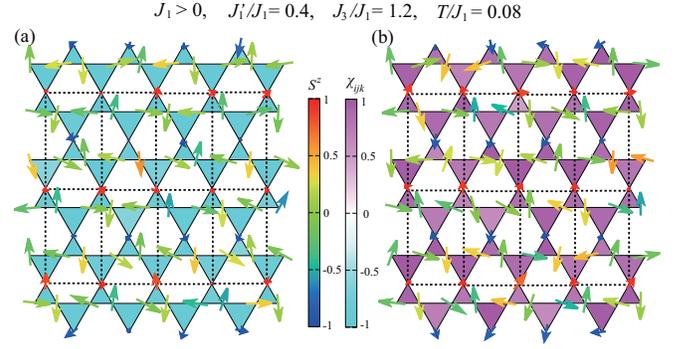}
\caption{MC snapshots of the noncoplanar chiral states taken at $T/J_1=0.08$ for $J_1'/J_1=0.4$ and $J_3/J_1=1.2$ with $J_1>0$. (a) SkX structure with negative total chirality $\chi^T$ and (b) anti-SkX structure with positive $\chi^T$. An arrow and its color represent the $S^xS^y$ and $S^z$ components of a spin, respectively, and the color of a triangle represents the local scalar chirality $\chi_{ijk}$ defined on each triangle. A unit cell of the skyrmion is indicated by a dotted rectangle. 
\label{fig:snapJ04} }
\end{center}
\end{figure}

In the uniform case of $J_1'/J_1=1$ [see greenish symbols in Fig. \ref{fig:Tdep} (a)], the collinearity $P$ develops below a transition temperature indicated by a sharp peak in $C$, whereas in the strongly breathing case of $J_1'/J_1=0.4$ [see blueish symbols in Fig. \ref{fig:Tdep} (a)], $P$ remains zero, but instead, the total chirality $\chi^{\rm T}$ develops. One can see from the spin snapshots shown in the top and bottom panels of Fig. \ref{fig:Tdep} (c) that the low-temperature phases for $J_1'/J_1=$1 and 0.4 are collinear and noncoplanar states, respectively, and that in the latter, three spins on different sublattices are orthogonal to one another. In the noncoplanar state, the spin correlation develops at the ordering vector of ${\bf q}={\bf Q}_1$, ${\bf Q}_2$, and ${\bf Q}_3$ [see Fig. \ref{fig:Tdep} (b)], and the topological number is $|n_{\rm sk}|=2$ [see the bottom panel of Fig. \ref{fig:Tdep} (a)], evidencing that the noncoplanar state is definitely the topological chiral order with the triple-${\bf Q}$ SkX structure. Here, $n_{\rm sk}=-2$ and $+2$ correspond to the SkX and anti-SkX with opposite $\chi^T$'s, respectively (see Fig. \ref{fig:snapJ04}), and the ordered state is the alternative of the two, similarly to other DM-free systems \cite{SkX_Okubo_12}.  

In the weakly breathing case of $J_1'/J_1=0.8$ [see the reddish symbols in Fig. \ref{fig:Tdep} (a)], $P$ exhibits two-step sudden drops, suggestive of successive first-order transitions. The higher-temperature and lower-temperature phases correspond to the collinear and noncoplanar states, respectively, and as one can see from the middle panel of Fig. \ref{fig:Tdep} (c), the intermediate phase is a coplanar state which turns out to be the one introduced in Fig. \ref{fig:structure} (b) (for details, see Supplemental Material \cite{Suppl}).   

Here, we address the nature of the phase transition. For the noncoplanar state, the chiral symmetry associated with the sign of $\chi^T$ is broken. In this chiral phase at $T\neq 0$, spins are not long-ranged-ordered and the chirality distribution is uniform (see Supplemental Material \cite{Suppl}), so that the translational symmetry of the underlying lattice is not broken. In this sense, strictly speaking, ''crystal'' of SkX is well-defined only at $T=0$ where spins are long-range-ordered. It should also be noted that a $\mathbb{Z}_2$-vortex transition \cite{Z2_Kawamura_84, Sqomega_Okubo_jpsj_10, Z2_AK_prl_20} may be additionally possible as spins are noncollinear in the low-temperature phase. In the transitions into the collinear and coplanar states, a local magnetization ${\bf m}_{\rm loc}$ and a local vector chirality $\mbox{\boldmath $\kappa$}_{\rm loc}$ defined on each triangle play important roles, respectively, where the manners of their spatial distributions, especially their quadratic correlations between the neighboring triangles, break the lattice translational symmetry (for details, see Supplemental Material \cite{Suppl}). 

Now that the emergence of the chiral order having the zero-field SkX structure is confirmed, we will next discuss its mechanism based on the mean-field approximation \cite{Reimers_MF}. We first introduce the Fourier transform ${\bf S}_i=\sum_{\bf q}{\bf S}^\alpha_{\bf q}\exp(i{\bf q}\cdot {\bf r}_i)$ with the site index $i=(\alpha, {\bf r}_i)$ and sublattice indices $\alpha=1, \, 2$, and 3 [see the inset of Fig. \ref{fig:structure} (a)], and pick up the ordering vectors of our interest ${\bf Q}_1$, ${\bf Q}_2$, and ${\bf Q}_3$. 
Then, the mean field $\langle {\bf S}^\alpha_{\bf q}\rangle$ can be expressed as
\begin{equation}\label{eq:S_q} 
\langle {\bf S}^\alpha_{\bf q}\rangle= \sum_{\mu=1}^3\big( [U^\alpha_{{\bf Q}_\mu}]^\ast\mbox{\boldmath $\Phi$}_{{\bf Q}_\mu} \delta_{{\bf q},{\bf Q}_\mu} + U^\alpha_{{\bf Q}_\mu} \mbox{\boldmath $\Phi$}_{{\bf Q}_\mu}^\ast \delta_{{\bf q},-{\bf Q}_\mu} \big),
\end{equation}
where $U^\alpha_{{\bf Q}_\mu}$ represents the $\alpha$th component of ${\bf U}_{{\bf Q}_\mu}$ which are given by 
\begin{eqnarray}\label{eq:Uq}
{\bf U}_{{\bf Q}_1} &=& ( 1, \, i\varepsilon, \, i\varepsilon)/\sqrt{1+2\varepsilon^2}, \nonumber\\
{\bf U}_{{\bf Q}_2} &=& (i \varepsilon, 1, \, i\varepsilon )/\sqrt{1+2\varepsilon^2},  \nonumber\\
{\bf U}_{{\bf Q}_3} &=& (i \varepsilon, \, i\varepsilon, \, 1)/\sqrt{1+2\varepsilon^2}
\end{eqnarray}
with dimensionless coefficient
\begin{eqnarray}
\varepsilon &=& \frac{ J_1-J_1'}{-\lambda_{{\bf Q}_\mu}+J_1+J_1'}, \\
\lambda_{{\bf Q}_\mu} &=& \frac{J_1+J_1'-4J_3-\sqrt{(J_1+J_1'+4J_3)^2+8(J_1-J_1')^2}}{2}. \nonumber
\end{eqnarray}
Note that $\varepsilon$ is zero only for $J'_1/J_1=1$, and gradually increases with decreasing $J'_1/J_1$.  

In the uniform case of $J'_1/J_1 =1$ and thereby $\varepsilon=0$, all the ${\bf U}_{{\bf Q}_\mu}$'s are orthogonal to one another, so that $\mbox{\boldmath $\Phi$}_{{\bf Q}_1}$, $\mbox{\boldmath $\Phi$}_{{\bf Q}_2}$, and $\mbox{\boldmath $\Phi$}_{{\bf Q}_3}$ correspond to the polarization vectors $\hat{P}_1$, $\hat{P}_2$, and $\hat{P}_3$, respectively. 
In other words, to make spins exist at each sublattice, the ordered state should be a triple-${\bf Q}$ state involving all the $\mbox{\boldmath $\Phi$}_{{\bf Q}_\mu}$'s, and thus, single-${\bf Q}$ and double-${\bf Q}$ states are not realized in the present system. Then, the Ginzburg-Landau (GL) free energy ${\cal F}_{\rm GL}$ is given by  ${\cal F}_{\rm GL}/(N/3) = f_2 + f_4 + \delta f_4$ with  
\begin{eqnarray}\label{eq:GL}
&& f_2 = \big[3T + \lambda_{{\bf Q}_\mu} \big] \sum_{\mu=1}^3 |\mbox{\boldmath $\Phi$}_{{\bf Q}_\mu}|^2, \nonumber\\
&& f_4 = \frac{9T}{20} \, A_1 \sum_{\mu=1}^3 \Big( |\mbox{\boldmath $\Phi$}_{{\bf Q}_\mu} \cdot \mbox{\boldmath $\Phi$}_{{\bf Q}_\mu}|^2 + 2 \big[\mbox{\boldmath $\Phi$}_{{\bf Q}_\mu} \cdot \mbox{\boldmath $\Phi$}^\ast_{{\bf Q}_\mu} \big]^2 \Big), \\
&& \delta f_4 = \frac{9T}{20}A_2 \sum_{\mu<\nu} \Big( |\mbox{\boldmath $\Phi$}_{{\bf Q}_\mu}|^2 |\mbox{\boldmath $\Phi$}_{{\bf Q}_\nu}|^2 + \sum_{\varepsilon_s=\pm 1} |\mbox{\boldmath $\Phi$}_{{\bf Q}_\mu} \cdot \mbox{\boldmath $\Phi$}_{\varepsilon_s{\bf Q}_\nu}|^2 \Big),  \nonumber
\end{eqnarray}
and the coefficients $A_1 = \frac{2(1+2\varepsilon^4)}{(1+2\varepsilon^2)^2}$ and $A_2 = \frac{8\varepsilon^2(2+\varepsilon^2)}{(1+2\varepsilon^2)^2} $.
Note that $\delta f_4$ is active only in the breathing case of $J'_1/J_1 \neq 1$, i.e., $\varepsilon \neq 0$.

The most stable spin configuration is obtained by minimizing ${\cal F}_{\rm GL}$ under the constraint ${\overline S}^2 = 2\sum_{\mu=1}^3 |\mbox{\boldmath $\Phi$}_{{\bf Q}_\mu}|^2$. 
Noting that in the uniform case, $\mbox{\boldmath $\Phi$}_{{\bf Q}_\mu}$ corresponds to the polarization vector ${\hat P}_\mu$, we assume $\mbox{\boldmath $\Phi$}_{{\bf Q}_\mu} = \frac{\overline S}{\sqrt{6}} e^{i \theta_\mu} {\hat P}_\mu$ also in the breathing case.
Then, one notices that the relative angles among ${\hat P}_1$,  ${\hat P}_2$, and ${\hat P}_3$ are determined only by $\delta f_4$ term which is calculated as
\begin{equation}\label{eq:delta_f4}
 \delta f_4 = \frac{T{\overline S}^4}{40} A_2 \Big[\frac{3}{2} + ({\hat P}_1\cdot{\hat P}_2)^2 + ({\hat P}_1\cdot{\hat P}_3)^2 + ({\hat P}_2\cdot{\hat P}_3)^2  \Big] .
\end{equation}
It is clear from Eq. (\ref{eq:delta_f4}) that in the breathing case of $\varepsilon \neq 0$ and thereby $A_2>0$, ${\hat P}_1$,  ${\hat P}_2$, and ${\hat P}_3$ become orthogonal to one another to lower the free energy $\delta f_4$, resulting in the noncoplanar SkX state. In the uniform case of $\varepsilon=0$ where the relative angles cannot be fixed because $\delta f_4=0$, thermal fluctuations favor the collinear state \cite{RMO-collinear_Grison_prb_20}. Also, the coplanar state is not stabilized at the level of the present GL expansion, so that higher-order terms or thermal fluctuations may be relevant to its stability.
 
\begin{figure}
\begin{center}
\includegraphics[width=\columnwidth]{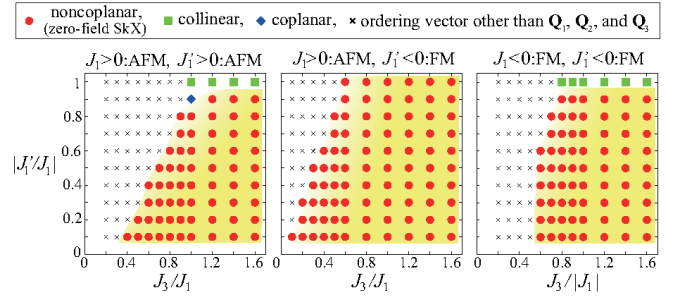}
\caption{Parameter dependence of the low-temperature spin structure at $T/|J_1|=0.01$. For large $J_3$, the chiral order with the noncoplanar SkX structure is realized irrespective of the sign of $J_1$ and $J_1'$. \label{fig:Paradep} }
\end{center}
\end{figure}

Since the above mean-field result is valid irrespective of the signs of $J_1$ and $J_1'$, the noncoplanar state should appear even in ferromagnetic $J_1$ and/or $J_1'$ as long as $J_1 \neq J_1'$ and thereby $\varepsilon \neq 0$. As shown in Fig. \ref{fig:Paradep}, this is actually the case: at the lowest temperature of our MC simulations, the noncoplanar SkX state is stable in the wide range of the parameter space except $J_1'/J_1=1$ at which the collinear state is realized. We have also checked that this chiral order is relatively robust against a magnetic field and a single-ion anisotropy as well as additional further NN interactions. 

For the present zero-field SkX, the sublattice degrees of freedom are fundamentally important because they make it possible to superpose three collinear chains, i.e., sinusoidal modulations, orthogonally, which is in contrast to the conventional in-field SkX where three helical modulations are superposed. Although in this work, the breathing bond-alternation is introduced to select the noncoplanar configuration from numerous energetically-degenerate superposition patterns, it might be substituted, on the uniform kagome lattice, by other driving forces such as a positive biquadratic spin interaction \cite{4sub_Akagi_prl_12, SkX_top2_Ozawa_prl_17, SkX-RKKY_Hayami_17}.

It is also useful to compare the present two-dimensional breathing-kagome system to the associated three-dimensional breathing-pyrochlore system. Although in both cases, the breathing bond-alternation commonly yields the noncoplanar topological spin textures, there is a significant difference in the properties of the ordered states. In the breathing-pyrochlore system, the hedgehog lattice, an alternating array of magnetic monopoles and antimonopoles of opposite topological charges, can be realized, but the total chirality summed over the whole system is completely canceled out at zero field \cite{hedgehog_AK_prb_21}. In contrast, in the present breathing-kagome system, the miniature discrete skyrmions of the same topological charges are condensed, and as a result, the total chirality is nonzero even at zero field, which suggests an interesting aspect of this two-dimensional magnet that in a metallic system, the Hall effect of chirality origin, the so-called topological Hall effect, may be observed even in the absence of an applied magnetic field.

Finally, we will address experimental implications of our results. Although several breathing-kagome magnets including Gd$_3$Ru$_4$Al$_{12}$ \cite{GdRuAl_Hirschberger_natcom_19} which hosts the in-field SkX have been reported so far \cite{DQVOF_Orain_prl_17, DQVOF_Clark_prl_13, LiAMoO_Haraguchi_prb_15, LiAMoO_Sharbaf_prl_18, PbOFCu_Zhang_ChemComm_20, YbNiGe_Takahashi_jpsj_20, FeSn_Tanaka_prb_20, CaCrO_Balz_natphys_16}, a magnetic order with commensurate ${\bf Q}_1$, ${\bf Q}_2$, and ${\bf Q}_3$ has not been observed. Nevertheless, in a metallic system, when the ordering vector is governed by the Ruderman-Kittel-Kasuya-Yosida interaction, one may have a chance to realize the commensurate order by tuning the Fermi level or the conduction electron density. If such a candidate material can be synthesized, the topological Hall effect may be observed at zero field as a distinctive signature of the zero-field chiral order with the triple-${\bf Q}$ miniature SkX structure.   

On the uniform kagome lattice, the 12-sublattice coplanar state shown in Fig. \ref{fig:structure} (b) has been reported in the magnetic insulator BaCu$_3$V$_2$O$_8$(OD)$_2$ \cite{CoplanarOct_Boldrin_prl_18} whose system parameters are consistent with our theory except the breathing lattice structure. If nominal lattice distortions, which might exist in this compound, yield nonequivalent $J_1$ and $J_1'$, the mechanism presented here could be applied to this class of magnets, pointing to the possibility of the chiral order in its family compounds.  

To conclude, we have theoretically demonstrated that the zero-field chiral order emerges in the form of the commensurate triple-${\bf Q}$ SkX in breathing-kagome antiferromagnets, where the third NN antiferromagnetic interaction along the bond direction $J_3$ and the breathing bond-alternation are essential. We believe that our study will promote the exploration of new classes of magnetic materials hosting topological spin textures.

\begin{acknowledgments}
The authors thank Y. Motome for useful discussions.
We are thankful to ISSP, the University of Tokyo and YITP, Kyoto University for providing us with CPU time. This work is supported by JSPS KAKENHI Grant Number JP17H06137 and JP21K03469.
\end{acknowledgments}

\hspace{2cm}
\pagebreak

\onecolumngrid
\hspace{2cm}
\begin{center}
\textbf{\large Supplemental Material for ''Emergent skyrmion-based chiral order in zero-field Heisenberg antiferromagnets on the breathing kagome lattice''}\\[.2cm]
\end{center}

\section{Monte Carlo results on the spin structures of the triple-${\bf Q}$ collinear and coplanar states}
\begin{figure}[b]
\begin{center}
\includegraphics[scale=0.95]{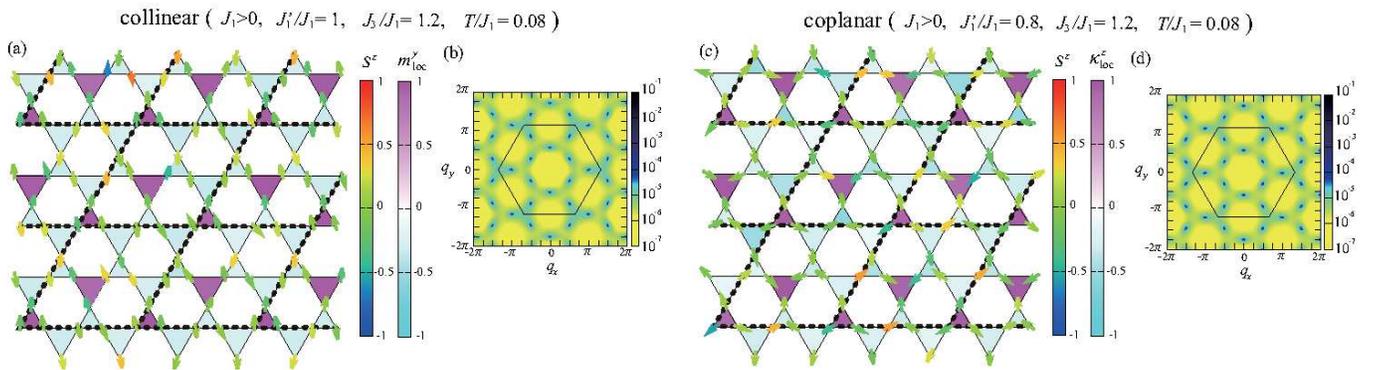}
\caption{MC results obtained in the triple-${\bf Q}$ collinear and coplanar states. (a) and (b) [(c) and (d)] spin snapshot and the associated spin structure factor $\langle | \frac{1}{N} \sum_i  {\bf S}_i \, e^{i {\bf q}\cdot{\bf r}_i}|^2 \rangle$ for the collinear (coplanar) state at $T/J_1=0.08$ for $J_1'/J_1=1$ ($J_1'/J_1=0.8$) and $J_3/J_1=1.2$ with $J_1>0$. Note that (a) and (c) are MC snapshots, whereas (b) and (d) are thermal-averaged quantities. In (a) and (c), an arrow and its color represent the $S^xS^y$ and $S^z$ components of a spin, respectively, and the color of a triangle represents (a) $y$ component of the local magnetization ${\bf m}_{\rm loc}$ and (c) $z$ component of the local vector chirality $\mbox{\boldmath $\kappa$}_{\rm loc}$ defined on each triangle. A magnetic unit cell is indicated by a dotted parallelogram. In (b) and (d), the wave number is measured in units of $1/a$ with with side length of each triangle $a$ and a hexagon indicates the border of the Brillouin zone.\label{fig:snapJ1-08} }
\end{center}
\end{figure}

Figure \ref{fig:snapJ1-08} shows the spin structures of the collinear and coplanar states obtained in the MC simulations. One can see from Figs. \ref{fig:snapJ1-08} (a) and (c) that the schematically-drawn spin structures introduced in Figs. 1 (a) and (b) in the main text are actually realized in the collinear and coplanar states, respectively. As shown in Figs. \ref{fig:snapJ1-08} (b) and (d), the spin structure factors for these states look the same as that for the triple-${\bf Q}$ noncoplanar state [see Fig. 2 (b) in the main text]: the spin correlation develops at the ordering wave vectors of  ${\bf Q}_1=\frac{\pi}{2a}(-1,-\frac{1}{\sqrt{3}})$,  ${\bf Q}_2=\frac{\pi}{2a}(1,-\frac{1}{\sqrt{3}})$, and ${\bf Q}_3=\frac{\pi}{2a}(0,\frac{2}{\sqrt{3}})$, evidencing that the collinear and coplanar states are the triple-${\bf Q}$ states. We note that throughout this paper, we ignore the bond-length difference in discussing, for example, ordering wave vectors, for simplicity, and take $a$ side length of each triangle, as the effect of the breathing bond-alternation has already been incorporated in the spin Hamiltonian in the form of the nonequivalent $J_1$ and $J_1'$. 

As we will explain in the next section, in the triple-${\bf Q}$ collinear and coplanar states, the local magnetization ${\bf m}_{\rm loc}$ and the local vector chirality $\mbox{\boldmath $\kappa$}_{\rm loc}$ defined on each triangle play important roles for a phase transition, respectively.

\section{order parameters for the triple-${\bf Q}$ collinear and coplanar states}
\begin{figure}
\begin{center}
\includegraphics[scale=0.75]{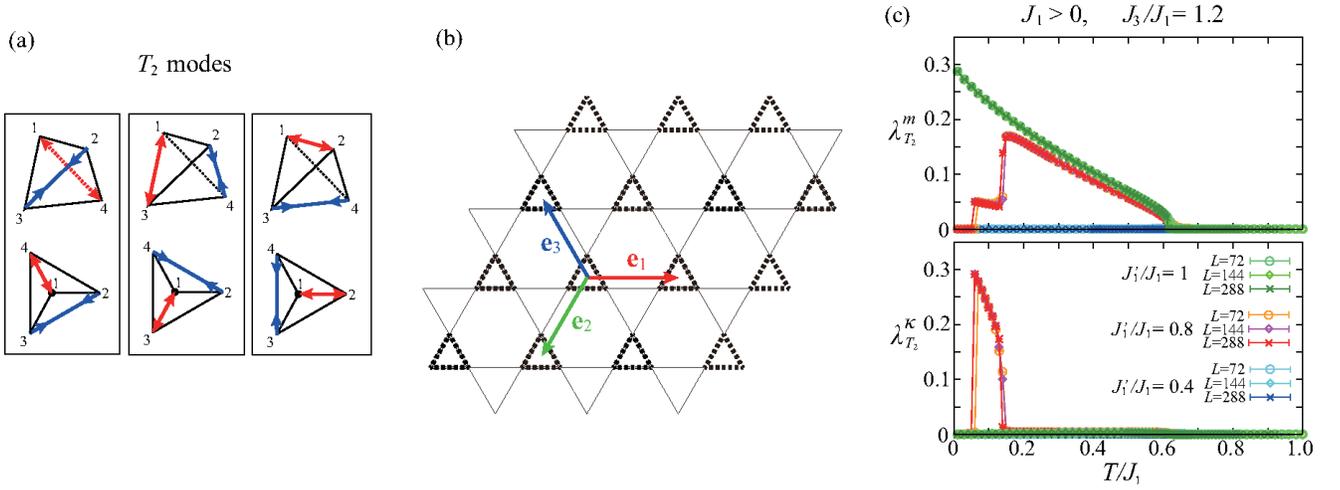}
\caption{Analogy between tetrahedral deformations and order parameters for the present breathing-kagome-lattice system. (a) Three $T_2$ modes of the tetrahedral deformations which elongate and compress opposing bonds of a tetrahedron. Upper panels are three-dimensional views and lower panels are the associated two-dimensional projections. (b) The breathing kagome lattice and the definitions of the vectors ${\bf e}_1$, ${\bf e}_2$, and ${\bf e}_3$. (c) Temperature dependence of the order parameter for the collinear state $\lambda^m_{T_2}$ (upper panel) and that for the coplanar state $\lambda^\kappa_{T_2}$ (lower panel) obtained in the MC simulations for $J_3/J_1=1.2$ with $J_1>0$, where greenish, reddish, and blueish colored symbols correspond to $J_1'/J_1=1$, $0.8$, and $0.4$, respectively. \label{fig:OP} }
\end{center}
\end{figure}
As is well known, in the present two-dimensional system, any long-range order of Heisenberg spin itself is not allowed at any finite temperature. In other words, spins are fluctuating even at low finite temperatures as in the paramagnetic phase. Nevertheless, as readily seen from the top panel of Fig. 1 (a) in the main text, the specific heat $C$ exhibits a divergent sharp peak suggestive of a phase transition from the paramagnetic phase into the triple-${\bf Q}$ collinear state. Noting that a discrete-symmetry breaking is possible even in two dimensions, here, we shall consider discrete degrees of freedom associated with the distributions of the local magnetization ${\bf m}_{\rm loc}$ and the local vector chirality $\mbox{\boldmath $\kappa$}_{\rm loc}$ which are defined as ${\bf m}_{\rm loc}=({\bf S}_i+{\bf S}_j+{\bf S}_k)/3$ and $\mbox{\boldmath $\kappa$}_{\rm loc}=\frac{2}{3\sqrt{3}}({\bf S}_i\times{\bf S}_j+{\bf S}_j \times {\bf S}_k +{\bf S}_k\times{\bf S}_i)$ with three spins on each triangle ${\bf S}_i$, ${\bf S}_j$, and ${\bf S}_k$.

As the triple-${\bf Q}$ states appearing in the present system are 12-sublattice states, there are four small triangles within the magnetic unit cell (see Fig. 1 in the main text). In the triple-${\bf Q}$ collinear and coplanar states, the distributions of ${\bf m}_{\rm loc}$ and  $\mbox{\boldmath $\kappa$}_{\rm loc}$ within the magnetic unit cell are characteristic. 

In the collinear case of Fig. \ref{fig:snapJ1-08} (a), among four small triangles within the magnetic unit cell, only one takes the three-up spin configuration (see the bottom left small triangle) and the rest three take the one-up and two-down  configurations. As one can see from the color contrast of the small triangles in  Fig. \ref{fig:snapJ1-08} (a), this means that among four ${\bf m}_{\rm loc}$'s on the small triangles, only one is not equivalent to the rest three. Such a situation is quite similar to a deformation of a tetrahedron whose four corners correspond to the four small triangles, so that we introduce the irreducible representations of the tetrahedral point group $T_d$. Among the six deformation modes, the three $T_2$ modes which can describe the trigonal distortions [see Fig. \ref{fig:OP} (a)] have bearing on the present system. Comparing the two-dimensional projections of the $T_2$ modes [see the lower panels of Fig. \ref{fig:OP} (a)] with the distributions of ${\bf m}_{\rm loc}$ on the small triangles [see Fig. \ref{fig:snapJ1-08} (a) together with Fig. \ref{fig:OP} (b)], one can find the analogy between the two. The three $T_2$ modes associated with ${\bf m}_{\rm loc}$, $\Lambda_{T_2,1}^m$, $\Lambda_{T_2,2}^m$, and $\Lambda_{T_2,3}^m$, are given by \cite{Bond_Shannon_10}

\begin{eqnarray}\label{eq:irreps_m}
\mbox{\boldmath $\Lambda$}_{T_2}^m({\bf R}_i)&=& \big(\Lambda_{T_2,1}^m({\bf R}_i), \Lambda_{T_2,2}^m({\bf R}_i), \Lambda_{T_2,3}^m({\bf R}_i) \big) \nonumber\\
\left( \begin{array}{c}
\Lambda_{T_2,1}^m ({\bf R}_i)\\
\Lambda_{T_2,2}^m ({\bf R}_i)\\
\Lambda_{T_2,3}^m ({\bf R}_i)
\end{array} \right) &=& \left( \begin{array}{cccccc}
0 & 0 & \frac{-1}{\sqrt{2}} & \frac{1}{\sqrt{2}} & 0 & 0 \\
0 &  \frac{-1}{\sqrt{2}} & 0 & 0 & \frac{1}{\sqrt{2}} & 0 \\
\frac{-1}{\sqrt{2}} & 0 & 0 & 0 & 0 &  \frac{1}{\sqrt{2}} 
\end{array} \right) \left( \begin{array} {c}
{\bf m}_{\rm loc} ({\bf R}_i) \cdot {\bf m}_{\rm loc}({\bf R}_i+{\bf e}_1)\\
{\bf m}_{\rm loc} ({\bf R}_i)\cdot {\bf m}_{\rm loc} ({\bf R}_i+{\bf e}_2)\\
{\bf m}_{\rm loc} ({\bf R}_i)\cdot {\bf m}_{\rm loc}({\bf R}_i+{\bf e}_3) \\
{\bf m}_{\rm loc} ({\bf R}_i+{\bf e}_1)\cdot {\bf m}_{\rm loc} ({\bf R}_i+{\bf e}_2)\\
{\bf m}_{\rm loc} ({\bf R}_i+{\bf e}_1) \cdot {\bf m}_{\rm loc}({\bf R}_i+{\bf e}_3) \\
{\bf m}_{\rm loc} ({\bf R}_i+{\bf e}_2)\cdot {\bf m}_{\rm loc} ({\bf R}_i+{\bf e}_3)
\end{array} \right) ,
\end{eqnarray}
where ${\bf R}_i$ is the center-of-mass position of a triangle and ${\bf e}_1$, ${\bf e}_2$, and ${\bf e}_3$ are vectors connecting neighboring small triangles [see Fig. \ref{fig:OP} (b)]. Then, we define the order parameter for the collinear state as
\begin{equation}\label{eq:OP_m}
\lambda^m_{T_2} = \Big\langle \Big| \frac{1}{L^2}\sum_{i \in \triangle} \mbox{\boldmath $\Lambda$}_{T_2}^m ({\bf R}_i) \Big|^2 \Big\rangle,
\end{equation}
where $\sum_{i \in \triangle}$ denotes the summation over all the small triangles. 

Now, we move on to the coplanar state. Comparing Figs. \ref{fig:snapJ1-08} (a) with (c), one notices that in the coplanar case of Fig. \ref{fig:snapJ1-08} (c), $\mbox{\boldmath $\kappa$}_{\rm loc}$'s are distributed in the same manner as that for ${\bf m}_{\rm loc}$ in the collinear case: within the magnetic unit cell, only the bottom left small triangle takes positive $\kappa^z_{\rm loc}$ and the rest three take negative values. Thus, similarly to the collinear case, we could define the order parameter in the coplanar case as
\begin{equation}\label{eq:OP_chi}
\lambda_{T_2}^\kappa = \Big\langle \Big| \frac{1}{L^2}\sum_{i \in \triangle} \mbox{\boldmath $\Lambda$}_{T_2}^\kappa ({\bf R}_i) \Big|^2 \Big\rangle,
\end{equation} 
where $\mbox{\boldmath $\Lambda$}_{T_2}^\kappa({\bf R}_i)$ is defined by simply replacing the local magnetization ${\bf m}_{\rm loc}({\bf R}_i)$ with the local vector chirality $\mbox{\boldmath $\kappa$}_{\rm loc}({\bf R}_i)$ in Eq. (\ref{eq:irreps_m}).

Figure \ref{fig:OP} (c) shows the temperature dependence of $\lambda_{T_2}^m$ and $\lambda_{T_2}^\kappa$ obtained for the same parameter set as that for Fig. 2 in the main text. As one can see from  Fig. \ref{fig:OP} (c), $\lambda_{T_2}^m$ and $\lambda_{T_2}^\kappa$ serve as the order parameters for the collinear and coplanar states, respectively, suggesting that the transitions into the collinear and coplanar state should accompany the lattice translational-symmetry breaking associated with the quadratic correlations of the local magnetization ${\bf m}_{\rm loc}$ and the local vector chirality $\mbox{\boldmath $\kappa$}_{\rm loc}$ between the neighboring triangles. We note that in the coplanar state, both $\lambda_{T_2}^m$ and $\lambda_{T_2}^\kappa$ are nonzero.

\begin{figure}
\begin{center}
\includegraphics[scale=0.95]{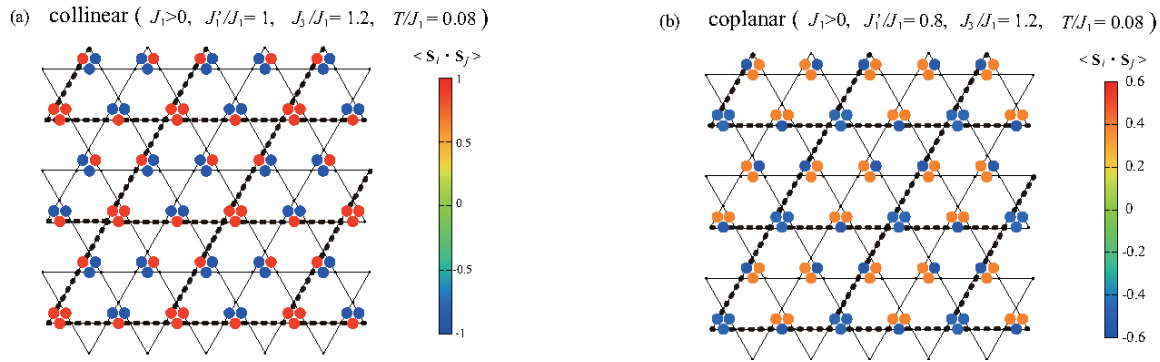}
\caption{The spatial distribution of the bond variable $\langle {\bf S}_i\cdot {\bf S}_j \rangle$ obtained in (a) the collinear state and (b) the coplanar state for the same system parameters as those in Fig. 1, where the color on a small-triangle bond represents the value of $\langle {\bf S}_i\cdot {\bf S}_j \rangle$. Note that (a) and (b) show the thermal-averaged value of $\langle {\bf S}_i\cdot {\bf S}_j \rangle$, in contrast to Figs. 1 (a) and (c) showing the MC snapshots. In both states (a) and (b), the lattice translational symmetry is spontaneously broken with a unit cell consisting of 12 spins indicated by a dotted parallelogram. \label{fig:bondsnapJ1-08} }
\end{center}
\end{figure}

The meaning of the lattice translational-symmetry breaking might be seen transparently in terms of the $O(3)$-symmetric bond variable $\langle {\bf S}_i\cdot {\bf S}_j \rangle$ which is quadratic in spins. One can see from the right-hand side of Eq. (\ref{eq:irreps_m}) that $\lambda_{T_2}^m$ is associated with $({\bf S}_i \cdot {\bf S}_j)$ because each ${\bf m}_{\rm loc}$ is linear in ${\bf S}_i$. In the collinear state shown in Fig. \ref{fig:snapJ1-08} (a), there exist 1-up and 2-down small triangles being arranged periodically. On each 1-up and 2-down triangle, the bond variable $({\bf S}_i \cdot {\bf S}_j)$ with two spins ${\bf S}_i$ and ${\bf S}_j$ at the both ends of the bond takes minus sign for an up-down pair, whereas positive sign for a down-down pair. Thus, the collinear state can be viewed as a periodic array of the positive and negative bond variables $({\bf S}_i \cdot {\bf S}_j)$, which can clearly be seen in the spatial distribution of the thermal-averaged value $\langle {\bf S}_i \cdot {\bf S}_j \rangle$ shown in Fig. \ref{fig:bondsnapJ1-08} (a). Such a situation is also the case for the coplanar state [see Fig. \ref{fig:bondsnapJ1-08} (b)]. As can be seen clearly from Figs. \ref{fig:bondsnapJ1-08} (a) and (b), these $O(3)$-symmetric bond variables which are quadratic in the original spin variables break the lattice translational symmetry, and the present collinear and coplanar orders might be regarded as classical analogues of quantum dimer orders or valence-bond crystals, although spins themselves are not long-range-ordered.

We emphasize that in the chiral phase with the noncoplanar SkX structure, both $\lambda_{T_2}^m$ and $\lambda_{T_2}^\kappa$ vanish [see Fig. \ref{fig:OP} (c)], so that the lattice translational symmetry is not broken not only in the spin sector but also in the $({\bf S}_i\cdot {\bf S}_j)$ and $({\bf S}_i \times {\bf S}_j)$ sectors. As will be explained below in Sec. III B, such a situation is also the case for the scalar chirality sector.

\section{notes on the structures of the triple-${\bf Q}$ noncollinear states}
\subsection{Spin configurations in the coplanar and noncoplanar states}
\begin{figure}
\begin{center}
\includegraphics[scale=0.48]{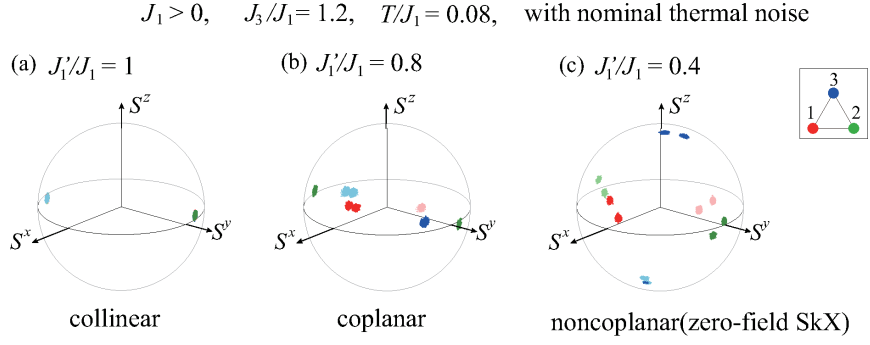}
\caption{Spin snapshots averaged over 200 MC sweeps to reduce thermal noises, where the system parameters are exactly the same as those for Fig. 2 (c) in the main text. The snapshots are taken in (a) the collinear state, (b) the coplanar state, and (c) the noncoplanar (zero-field SkX) state. Red, green, blue dots represent spins at the three corners of the small triangle shown in the inset. \label{fig:snap_average} }
\end{center}
\end{figure}

In Fig. 2 (c) in the main text, we show the spin snapshots mapped onto a unit sphere in the triple-${\bf Q}$ collinear, coplanar, and noncoplanar states, where the difference among the three consists in the relative angles among the three polarization vectors $\hat{P}_1$, $\hat{P}_2$, and $\hat{P}_3$. In particular, in the noncoplanar SkX state, $\hat{P}_1$, $\hat{P}_2$, and $\hat{P}_3$ are almost orthogonal to one another, although spins on each sublattice [see the same colored dots in Fig. 2 (c) in the main text] are scattered because of the thermal noise. In order to look into further details of the spin structures, we shall take a short-time average of the spin snapshots to reduce the thermal noise. 

Figure \ref{fig:snap_average} shows a 200-MC-sweeps averaged version of the spin snapshots shown in Fig. 2 (c) in the main text. One can see from Fig. \ref{fig:snap_average} that the polarization vector of each sublattice splits into two in the coplanar and noncoplanar states, while not in the collinear state. As we will address below, such a splitting is generic to the triple-${\bf Q}$ noncollinear states on the breathing kagome lattice, and its origin can be understood within the mean-field approximation.

Since the Fourier component of spin $\langle {\bf S}^\alpha_{\bf q}\rangle$ is obtained with the combined use of $\mbox{\boldmath $\Phi$}_{{\bf Q}_\mu} = \frac{\overline S}{\sqrt{6}} e^{i \theta_\mu} {\hat P}_\mu$ and Eqs. (2) and (3) in the main text, spins belonging to sublattice $\alpha$, ${\bf S}_j^\alpha$'s, are given by
\begin{eqnarray}\label{eq:spin_sublattice}
{\bf S}_j^1 &=& c_S\Big[ \hat{P}_1 \cos({\bf Q}_1\cdot{\bf r}_j + \theta_1) +\varepsilon \big\{\hat{P}_2 \sin({\bf Q}_2\cdot{\bf r}_j + \theta_2) + \hat{P}_3 \sin({\bf Q}_3\cdot{\bf r}_j + \theta_3) \big\} \Big], \nonumber\\
{\bf S}_j^2 &=& c_S\Big[ \hat{P}_2 \cos({\bf Q}_2\cdot{\bf r}_j + \theta_2)  +\varepsilon \big\{ \hat{P}_1 \sin({\bf Q}_1\cdot{\bf r}_j + \theta_1) + \hat{P}_3 \sin({\bf Q}_3\cdot{\bf r}_j + \theta_3)  \big\} \Big], \nonumber\\
{\bf S}_j^3 &=& c_S \Big[ \hat{P}_3 \cos({\bf Q}_3\cdot{\bf r}_j + \theta_3) +\varepsilon \big\{\hat{P}_1 \sin({\bf Q}_1\cdot{\bf r}_j + \theta_1) + \hat{P}_2 \sin({\bf Q}_2\cdot{\bf r}_j + \theta_2) \big\} \Big], \nonumber\\
c_S &=& \frac{{\overline S}}{\sqrt{1+2\varepsilon^2}}\sqrt{\frac{2}{3}}.  
\end{eqnarray}
As long as $\hat{P}_\mu$'s are not collinear like in the cases of the coplanar and noncoplanar structures, spins on each sublattice cannot be collinear because of the existence of $\varepsilon$, which can clearly be seen from the following calculation. In the triple-${\bf Q}$ state, the magnetic unit cell contains four small triangles. Here, we label these four as triangles I, II, III, and IV [see Fig. \ref{fig:Schi_dist} (a)]. Suppose that the positions of sublattice 1 on triangles I, II, III, and IV be ${\bf r}_{\rm I}=(0,0)$, ${\bf r}_{\rm II}=a(2,0)$, ${\bf r}_{\rm III}=a(1,\sqrt{3})$, and ${\bf r}_{\rm IV}=a(3,\sqrt{3})$, respectively. Then, the three sublattice spins on the four small triangles I, II, III, and IV are expressed as
\begin{eqnarray}
&& \left\{ \begin{array}{cc}
{\bf S}^1_{\rm I} = c_S\big[ \hat{P}_1\cos\theta_1+\varepsilon (\hat{P}_2\sin\theta_2+\hat{P}_3\sin\theta_3)\big], & {\bf S}^1_{\rm II} = c_S\big[ -\hat{P}_1\cos\theta_1-\varepsilon (\hat{P}_2\sin\theta_2-\hat{P}_3\sin\theta_3)\big], \nonumber\\
{\bf S}^1_{\rm III} = c_S\big[ -\hat{P}_1\cos\theta_1+\varepsilon (\hat{P}_2\sin\theta_2-\hat{P}_3\sin\theta_3)\big], & {\bf S}^1_{\rm IV} = c_S\big[ \hat{P}_1\cos\theta_1-\varepsilon (\hat{P}_2\sin\theta_2+\hat{P}_3\sin\theta_3)\big],
\end{array} \right . \nonumber\\
&& \left\{ \begin{array}{cc}
{\bf S}^2_{\rm I} = c_S\big[ -\hat{P}_2\sin\theta_2-\varepsilon (\hat{P}_1\cos\theta_1-\hat{P}_3\sin\theta_3)\big], & {\bf S}^2_{\rm II} = c_S\big[\hat{P}_2\sin\theta_2+\varepsilon (\hat{P}_1\cos\theta_1+\hat{P}_3\sin\theta_3)\big], \nonumber\\
{\bf S}^2_{\rm III} = c_S\big[ -\hat{P}_2\sin\theta_2+\varepsilon (\hat{P}_1\cos\theta_1-\hat{P}_3\sin\theta_3)\big], & {\bf S}^2_{\rm IV} = c_S\big[ \hat{P}_2\sin\theta_2-\varepsilon (\hat{P}_1\cos\theta_1+\hat{P}_3\sin\theta_3)\big],
\end{array} \right . \nonumber\\
&& \left\{ \begin{array}{cc}
{\bf S}^3_{\rm I} = c_S\big[ -\hat{P}_3\sin\theta_3-\varepsilon (\hat{P}_1\cos\theta_1-\hat{P}_2\sin\theta_2)\big], & {\bf S}^3_{\rm II} = c_S\big[ -\hat{P}_3\sin\theta_3+\varepsilon (\hat{P}_1\cos\theta_1-\hat{P}_2\sin\theta_2)\big], \nonumber\\
{\bf S}^3_{\rm III} = c_S\big[ \hat{P}_3\sin\theta_3+\varepsilon (\hat{P}_1\cos\theta_1+\hat{P}_2\sin\theta_2)\big], & {\bf S}^3_{\rm IV} = c_S\big[ \hat{P}_3\sin\theta_3-\varepsilon (\hat{P}_1\cos\theta_1+\hat{P}_2\sin\theta_2)\big],
\end{array} \right . 
\end{eqnarray}   
so that in the coplanar and noncoplanar states, all the four spins on each sublattice within the magnetic unit cell orient in the different directions, exhibiting the two-spots pair when the spins are mapped onto the sphere.
Since $\varepsilon$ is nonzero in the breathing case, such a splitting is inherent to the triple-${\bf Q}$ noncollinear states on the breathing kagome lattice. 

In Figs. 1 (c) and (d) in the main text, we illustrate, for simplicity, the real-space structure of the noncoplanar SkX state such that spins belonging to each sublattice are perfectly collinear. But, strictly speaking, this is not the case from the reason explained above. 
Nevertheless, as readily seen from the lowest panel of Fig. 2. (a) in the main text, the skyrmion number is definitely $n_{\rm sk}=\pm 2$ even in this slightly-disturbed situation. 
Thus, we could conclude that $\varepsilon$ is important for the emergence of the noncoplanar SkX state as explained in the main text, but the resultant slight deviation from the perfectly collinear $\uparrow\downarrow\uparrow\downarrow$ structure is not.

\subsection{Scalar-chirality distribution in the noncoplanar state}
\begin{figure}
\begin{center}
\includegraphics[scale=0.75]{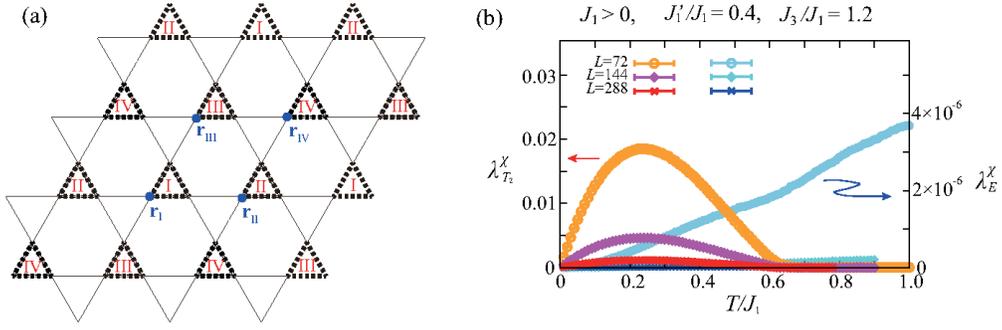}
\caption{(a) Definition of the four small triangles I, II, III, and IV within the 12-sublattice magnetic unit cell (see the text in Sec. III in Supplementary Material). (b) Temperature dependence of the order parameters measuring the nonequivalency of the four scalar schiralities $\chi_{\rm I}$, $\chi_{\rm II}$, $\chi_{\rm III}$, and $\chi_{\rm IV}$, $\lambda^\chi_{T_2}$ (reddish symbols) and $\lambda^\chi_{E}$ (blueish symbols), obtained in the MC simulations for $J_1'/J_1=1$ and $J_3/J_1=1.2$ with $J_1>0$. \label{fig:Schi_dist} }
\end{center}
\end{figure}
As addressed in the main text, at the transition from the paramagnetic phase into the noncoplanar SkX state, the $\mathbb{Z}_2$ chiral symmetry associated with the sign of the total scalar chirality is broken spontaneously although spins themselves remain disordered, namely, the so-called spin-chirality decoupling \cite{SG_Kawamura} occurs. In Sec. II, we show that in this chiral phase, the lattice translational symmetry associated with the quadratic correlations of the spins (the vector chiralities) between the neighboring sites (triangles) are spontaneously broken. Then, the natural question is whether the spatial distribution of the scalar spin chirality $\chi_{ijk}={\bf S}_i \cdot ({\bf S}_j \times {\bf S}_k)$, which is relevant to this chiral phase, breaks the lattice translational symmetry or not. Since as explained in Sec. III A, instantaneous or short-time spin configurations on the four small triangles I, II, III, and IV within the magnetic unit cell are not equivalent to one another in the noncoplanar chiral phase, one might expect that $\chi_{ijk}$'s defined on the four small triangles are also not equivalent, but this is not the case. To see this, we define an order parameter measuring the nonequivalency of the scalar chirality in the same manner as that in Sec. II. We map the four scalar chiralities $\chi_{\rm I}$, $\chi_{\rm II}$, $\chi_{\rm III}$, and $\chi_{\rm IV}$ averaged over each of four types of triangles I, II, III, and IV in the whole system onto the four corners of a tetrahedron. Then, the nonequivalency of the four scalar chiralities is characterized by the tetrahedron deformations except the uniform $A_1$ mode, namely, the three $T_2$ and two $E$ modes, which are given by \cite{Bond_Shannon_10}
\begin{equation}\label{eq:irreps_chi}
\left( \begin{array}{c}
\Lambda_{T_2,1}^{\chi} \\
\Lambda_{T_2,2}^{\chi} \\
\Lambda_{T_2,3}^{\chi} \\
\Lambda_{E,1}^{\chi} \\
\Lambda_{E,2}^{\chi}
\end{array} \right) = \left( \begin{array}{cccccc}
0 & 0 & \frac{-1}{\sqrt{2}} & \frac{1}{\sqrt{2}} & 0 & 0 \\
0 &  \frac{-1}{\sqrt{2}} & 0 & 0 & \frac{1}{\sqrt{2}} & 0 \\
\frac{-1}{\sqrt{2}} & 0 & 0 & 0 & 0 &  \frac{1}{\sqrt{2}} \\
\frac{1}{\sqrt{3}} & \frac{-1}{2\sqrt{3}} & \frac{-1}{2\sqrt{3}} & \frac{-1}{2\sqrt{3}} & \frac{-1}{2\sqrt{3}} & \frac{1}{\sqrt{3}} \\
0 & \frac{1}{2} & \frac{-1}{2} & \frac{-1}{2} & \frac{1}{2} & 0
\end{array} \right) \left( \begin{array} {c}
\chi_{\rm I} \, \chi_{\rm II}\\
\chi_{\rm I} \, \chi_{\rm III}\\
\chi_{\rm I} \, \chi_{\rm IV} \\
\chi_{\rm II} \, \chi_{\rm III}\\
\chi_{\rm II} \, \chi_{\rm IV} \\
\chi_{\rm III} \, \chi_{\rm IV}
\end{array} \right)  
\end{equation}
with $\chi_{\alpha} = \frac{1}{L^2/4}\sum_{ijk \in {\rm triangle}\, \alpha} \chi_{ijk}$ ($\alpha=$ I, II, III, and IV) .
The associated order parameters are defined by
\begin{equation}
\lambda_{T_2}^\chi = \Big\langle \big( \Lambda_{T_2,1}^{\chi} \big)^2+\big(\Lambda_{T_2,2}^{\chi}\big)^2+\big(\Lambda_{T_2,3}^{\chi}\big)^2 \Big\rangle, \qquad \lambda_{E}^\chi = \Big\langle \big(\Lambda_{E,1}^{\chi}\big)^2+\big(\Lambda_{E,2}^{\chi}\big)^2 \Big\rangle.
\end{equation}
Figure \ref{fig:Schi_dist} (b) shows the temperature dependence of $\lambda_{T_2}^\chi$ and $\lambda_{E}^\chi$ for $J_1'/J_1=0.4$, $J_3/J_1=1.2$, and $J_1>0$, the same parameter set as that for the strongly breathing case in Fig. 2 in the main text, where the low-temperature phase below $T/J_1=0.61$ is the chiral phase with the noncoplanar SkX structure. One can see that both $\lambda_{T_2}^\chi$ and $\lambda_{E}^\chi$ are suppressed with increasing the system size $L$, eventually vanishing in the thermodynamic limit of $L\rightarrow \infty$. This suggests that the scalar chiralities $\chi_{\rm I}$, $\chi_{\rm II}$, $\chi_{\rm III}$, and $\chi_{\rm IV}$ defined on the four small triangles within the magnetic unit cell are equivalent to one another. 

Looking at the center of the skyrmion in Fig. 1 (d) in the main text, one notices that there are four different types of triangles including the small one [triangles 1-4 in Fig. 1 (d) in the main text]. As explained above, in the case of the small triangle, $\chi_{\rm I}$, $\chi_{\rm II}$, $\chi_{\rm III}$, and $\chi_{\rm IV}$ are equivalent. Such a situation is also the case for the chiralities defined on each of rest three types of triangles, so that the chirality distribution is uniform, keeping the translational symmetry of the underlying lattice.

\end{document}